\documentclass[aps,prl,toolkits,twocolumn]{revtex4}

\usepackage{epsfig,amssymb,amsfonts}

\makeatletter

\renewcommand{\@makefntext}[1]
{\parindent=1em\noindent\hbox to 1.8em{\hss$^{\@thefnmark}$}#1}
\renewcommand{\@footnotemark}{\hbox{\mathsurround=0pt$^{\@thefnmark}$}}

\makeatother

\begin{document}
\title{Is a consistent holographic description of excited hadrons with fixed L possible?}

\maketitle

In  an interesting recent Letter \cite{B}  de Teramond 
and Brodsky suggest a holographic
description of excited hadrons within the bottom-up  AdS/QCD. A key
statement is that a holographic description of excited hadrons with the
fixed (conserved) orbital angular momentum $L$ and with the principal quantum number $\sim n_r + L$
is found.  Here we prove that the existing
holographic constructions that rely on the standard matching condition at the ultraviolet
boundary $z=0$ do not allow the existence of a conserved orbital angular momentum $L$ of quarks
in each meson in 3+1 dimensions and that consequently the de Teramond - Brodsky
construction  violates either unitarity or  chiral symmetry
in the ultraviolet.

The heart of the holographic description of hadrons is a matching of the  QCD operators
composed of free quarks and gluons with the required quantum numbers  at the ultraviolet
boundary $z=0$ with their duals in the 5-dim AdS space. Since there is no chiral symmetry
breaking in the ultraviolet these QCD composite operators transform according to some representations
of the chiral $SU(2)_L \times SU(2)_R$ (we consider for simplicity the chiral limit of two
flavor QCD). The chiral classification of these operators can be found in Refs. \cite{CJ,G1}.
There is a unitary transformation from the quark-antiquark chiral basis to the standard
nonrelativistic $^{2S+1}L_J$ basis \cite{GN}.  It represents a straightforward generalization
of the unitary transformation from the two-particle relativistic helicity basis to the
nonrelativistic basis \cite{L}. Then, according to this transformation, it is obvious that
it is not possible in general to attach to a given quark-antiquark composite operator 
a dual with quantum numbers that include standard nonrelativistic angular momenta
$^{2S+1}L_J$ in 3+1 dimensions with fixed $L$. 

In order to make this point clear, consider as an example,
a holographic description of $\rho$-mesons. Typically
the   $\rho$-mesons are described as  holographic duals of the vector current $\bar q \gamma^\mu \vec \tau q$
 in QCD. Not only this operator, however, can create  
$\rho$-mesons from the vacuum in QCD. There is another local operator with the same dimension that also
creates $\rho$-mesons, it is the pseudotensor current  $\bar q \sigma^{0i} \vec \tau q$. It is
well established on the lattice that the physical $\rho$-meson couples equally well  
to both currents \cite{lattice}. A key point is that the vector and the pseudotensor currents have
radically different chiral transformation properties. While the vector current transforms as
$(1,0) + (0,1)$ of the $SU(2)_L \times SU(2)_R$, the pseudotensor current belongs to the $(1/2,1/2)_b$
representation \cite{CJ,G1}. Then a unitary transformation from the  quark-antiquark vector
and pseudotensor currents with the $\rho$-meson quantum numbers to the 
standard nonrelativistic $^{2S+1}L_J$ basis is given as \cite{GN}:

\begin{eqnarray}
\displaystyle |(0,1)+(1,0);1 ~ 1^{--}\rangle&=
&\sqrt{\frac23}|1;{}^3S_1\rangle+\sqrt{\frac13}|1;{}^3D_1\rangle,\nonumber\\
\displaystyle |(1/2,1/2)_b;1 ~ 1^{--}\rangle&=
&\sqrt{\frac13}|1;{}^3S_1\rangle-\sqrt{\frac23}|1;{}^3D_1\rangle.\nonumber
\end{eqnarray}

\noindent
Consequently each of the currents represents a superposition of two different
vectors with different $L$.
Hence it is not possible to construct a dual to a composite   $\rho$--meson 
operator with  fixed chiral transformation properties that would respond to 
one of the fixed
$^3S_1$ or $^3D_1$ sets of angular momenta. Of course, it is possible to construct a linear
combination of the vector and pseudotensor currents that have a definite fixed $L$.
But then such an operator will be a mixture of different chiral representations
and would correspond to a broken chiral symmetry. That would contradict  a basic
element of a holographic construction that at the ultraviolet border the chiral symmetry cannot be broken. 
This argument is  general, it depends only on the identification of $S$ and $L$ in the
holographic theory with their standard 3+1 dimensional meanings;
it is valid not only for the local $J=1$
operators, but also for the quark-antiquark higher spin mesons.

This proves a kind of  no-go theorem that it is impossible in general to construct a 
holographic description of {\it each} meson (with the quark-antiquark valence degree of fredom
only)  with fixed (conserved) quantum number $L$  in 3+1 dimensions for each meson and that 
would satisfy at the same time manifest chiral symmetry in the ultraviolet  and would not violate
unitarity.

The author thanks Stan Brodsky, Tom Cohen and Guy de Teramond
for discussions and acknowledges support of the Austrian Science Fund through
Grant No. P19168-N16.

{ Leonid Ya. Glozman} \\ \\
 {Institute for
Physics, University of Graz, Universit\"atsplatz 5, A-8010 Graz,
Austria}

\end{document}